\documentclass[aip, jcp, reprint, groupedaddress, showpacs, superscriptaddress, amsmath, amssymb]{revtex4-1} 
\usepackage[colorlinks,allcolors=blue]{hyperref}
\usepackage{graphicx}
\usepackage{dcolumn}
\usepackage{epstopdf}
\usepackage{bm}
\usepackage{braket}

\newcommand\fitwidth{0.5\textwidth}

\newcommand{\beginsupplement}{%
	\setcounter{equation}{0}
	\renewcommand{\theequation}{A\arabic{equation}}%
}

\begin{document}
	\title{Momentum-resolved TDDFT algorithm in atomic basis for real
		time tracking of electronic excitation}
	\author{Chao Lian}
	\author{Shi-Qi Hu}
	\author{Meng-Xue Guan}
	\affiliation{Beijing National Laboratory for Condensed Matter
		Physics and Institute of Physics, Chinese Academy of Sciences,
		Beijing, 100190, P. R. China}
	
	\author{Sheng Meng}
	\email{smeng@iphy.ac.cn}
	\affiliation{Beijing National Laboratory for Condensed Matter Physics and Institute of Physics, Chinese Academy of Sciences,
		Beijing, 100190, P. R. China}
	\affiliation{Collaborative Innovation Center of Quantum
		Matter, Beijing, 100190, P. R. China}
	
	\date{\today}

    \begin{abstract}
Ultrafast electronic dynamics in solids lies at the core of modern condensed matter and materials physics. To build up a practical ab initio method for studying solids under photoexcitation,
we develop a momentum-resolved real-time time dependent density functional theory (rt-TDDFT)
algorithm using numerical atomic basis, together with the implementation of both the length and
vector gauge of the electromagnetic field. When applied to simulate elementary excitations in
two-dimensional materials such as graphene, different excitation modes, only distinguishable in
momentum space, are observed. The momentum-resolved rt-TDDFT is important and computationally efficient for the study of ultrafast dynamics in extended systems.
    \end{abstract}

    \maketitle
    \section{Introduction}
    Real-time (rt) time dependent density functional theory
    (TDDFT) is an efficient \textit{ab initio} method to study electron dynamics in complex electron-nuclear systems in both
    the ground state and excited state. Compared with other
    widely used approaches such as frequency domain TDDFT,
    quasi-particle GW, and Bethe-Salpeter equations, rt-TDDFT
    has two major advantages: (i) Time-dependent Kohn-Sham
    (TDKS) equations in rt-TDDFT include all nonlinear effects
    and are intrinsically non-perturbative, making rt-TDDFT a
    better tool to describe materials in a strong field and (ii)
    rt-TDDFT directly provides complete information on real
    time evolution of electronic wavefunctions together with ionic
    movements, presenting a unique way for real-time tracking ultrafast dynamics and complex phenomena far from equilibrium. Thus, rt-TDDFT is a natural choice for the exploration of strong field physics and ultrafast phenomena. Motivated by the rapid developments in ultrafast experimental
    techniques, e.g., attosecond based spectroscopy~\cite{RevModPhys.81.163}, ultrastrong laser sources~\cite{RevModPhys.90.021002} and free electron X-ray lasers~\cite{RevModPhys.88.015006}, rt-TDDFT
    is drawing more and more attention as a method to simulate ultrafast phenomena in the current line of research
    frontiers.
    
Nevertheless, rt-TDDFT is not widely used as the method
of choice in the literature, being much less popular than other
density functional theory (DFT) based approaches such as
$\Delta$SCF, DFT+U, frequency-domain TDDFT, etc. 
Thus, numerical atomic orbitals (NAO) have been
a common choice to dramatically reduce computation cost for
simulating complex materials and have been widely used in DFT codes such as SIESTA~\cite{Soler2002, Ordejon1996} and OpenMX~\cite{PhysRevB.67.155108} and rt-TDDFT implementations by A. Tsolakidis~\cite{Tsolakidis2002} and X. Li~\cite{Li2005a,Isborn2007}. The biggest advantage of using NAO is the extremely small computational cost. To describe a system with $N_a$ atoms, only about $10 \times N_a$ NAOs are required, while $10^3 - 10^4 \times N_a$ real space grids or plane waves have to be invoked. In addition, with a relatively small real-space cutoff for NAOs, the order-$N$ linear
scaling with respect to system size can be achieved. Since a
major difficulty in developing rt-TDDFT is its extreme time
consumption due to the use of ultrasmall time step (on the order of $\sim$1 attosecond), NAO based $\mathbf{k}$-resolved rt-TDDFT is very promising for simulating realistic condensed matter systems, complex materials, and interfaces with a long simulation time.
    
    Most previous rt-TDDFT investigations focus on the
    photoabsorption and related properties of finite-size zero-dimensional (0D) systems (atoms/molecules/nanoparticles)
    including optical spectra,~\cite{Yamamoto2006, Qian2006, Tong2001, Tong1998, Tong1997, Mirzaei2010, Nobusada2004, Ganeev2015, Lopata2012, Fernando2015, Tussupbayev2015,  Lopata2013, Raghunathan2012, Williams-Young2016, Bruner2016a, Provorse2015, Fischer2015, Repisky2015, Lopata2011, Nguyen2015, Zheng2016}, excited state dynamics~\cite{Donati2016,Petrone2014,Chapman2011}, solvation effect~\cite{Donati2017a,Ding2015,Chapman2013,Nguyen2012,Liang2012}, relativistic effect variationally~\cite{Kasper2018,Goings2016}, photochemical stability~\cite{Haruyama2012, Haruyama2012a, Hu2013a, Silaeva2015, Yan2016}, and recently plasmonic excitations~\cite{Ding2014, Donati2017, Donati2018, Manjavacas2014, Barbry2015, Townsend2011a, Ma2015a, Yan2011, Song2012, Yan2007b, Gao2011, Gao2015c, Song2011}. In 0D systems, only single $\Gamma$ point is needed in
    the reciprocal space sampling. Thus, the $\Gamma$-only algorithm
    is overwhelming, as commonly implemented and used in
    the majority of rt-TDDFT simulations. However, to study
    photoexcitation and electronic dynamics in extended systems, $\Gamma$-only k-point sampling is insufficient and momentum-resolved ($\mathbf{k}$-resolved) sampling in the reciprocal space is required. 

    An important advantage of using $\mathbf{k}$-resolved rt-TDDFT is computational efficiency. With ${\Gamma}$-only TDDFT, to get the
    accurate charge density and ionic forces, an extraordinary large supercell has to be invoked. Many previous studies on extended systems belong to this scenario,~\cite{Miyamoto2006, Miyamoto2007a, Krasheninnikov2007, Miyamoto2007, Zhang2009, Miyamoto2010b, Zhang2012a, Zhang2012b, Miyamoto2013} including our
    recent studies on ultrafast electron-hole dynamics in dye-sensitized solar cells,~\cite{Meng2010, Meng2008a, Ma2014, Ma2013, Jiao2011a, Jiao2013a} charge separation in van der Waals heterojunctions,~\cite{Zhang2017a} and nonthermal melting of silicon.~\cite{Lian2016} Using $\mathbf{k}$-resolved algorithms, and at the same accuracy level, the supercell size as well as the computational cost, can be largely reduced, as will be demonstrated later. 
    
    Besides technical advantages, $\mathbf{k}$-resolved algorithm introduces the important $\mathbf{k}$-space resolution and a new degree
    of freedom, which is essential to describe key quantities and important physics in condensed matter materials such as time-dependent band structures, quasiparticles, and valley dynamics. Only rt-TDDFT with $\mathbf{k}$-resolved sampling can provide essential information concerning the real time evolution of
    material properties.
    
    Although $\mathbf{k}$-resolved rt-TDDFT algorithms have been implemented by several groups~\cite{Bertsch2000, Marques2003, Castro2006, Andrade2015} and applied for both semiconductors,~\cite{Sato2014, Sato2015, Otobe2016, Yabana2012, Shinohara2010a, Shinohara2010, Otobe2009, Otobe2008, Wachter2014} and metals,~\cite{Krieger2015, Elliott2016, Schleife2012,Yost2017} these implementations employ either real space grids or planewaves as basis sets. With
    a much smaller basis set, the implementation of $\mathbf{k}$-resolved rt-TDDFT algorithms with NAO basis has advantages in efficiency. To take the advantages of NAOs, a new framework and a more complicated implementation of rt-TDDFT are required.
    
    In this work, we strive to tackle the major challenges mentioned above in NAO-based rt-TDDFT. We have successfully developed the $\mathbf{k}$-resolved rt-TDDFT algorithm based on local atomic basis sets using numerical atomic orbitals.
    Both the length and vector gauge of the electromagnetic
    field have been implemented. This approach enables rt-TDDFT calculations of solids and surfaces using rather
    simple unit cells, reducing computational cost by several orders of magnitudes. Moreover, momentum-resolved electron dynamics in the excited states can be tackled by this approach. For instance, $\mathbf{k}$ selective photoexcitations in
    graphene are demonstrated here, where three distinct photoexcitation modes located at different $\mathbf{k}$points in the reciprocal space are induced upon laser illumination. This kind of $\mathbf{k}$-dependent electronic dynamics is ubiquitous in extended systems such as periodic solids and interfaces. Therefore, we expect highly efficient $\mathbf{k}$-resolved rt-TDDFT algorithms
    employing local bases be an important development and will be widely used in first-principles simulations of ultrafast phenomena under a strong field and optimal control of quantum materials.

    \section{Methodology}
\begin{figure}
    	\centering
    	\includegraphics[width=\fitwidth]{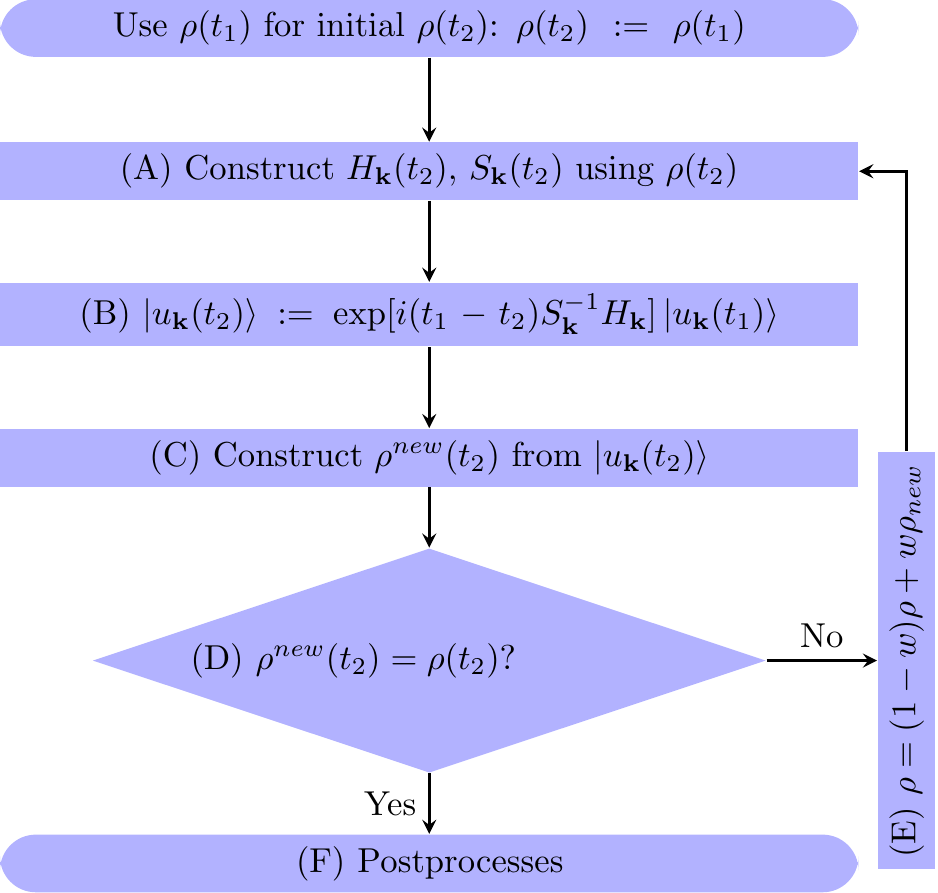}
    	\caption{\label{flowchart} Flowchart of $\mathbf{k}$-resolved rt-TDDFT algorithm. Here $S_\mathbf{k}$ is the overlap matrix, $H_\mathbf{k}$ is the Hamiltonian matrix, and $\ket{u_{n\mathbf{k}}}$ is the periodic part of TDKS orbitals at momentum $\mathbf{k}$.}
    \end{figure}

    The main framework of $\mathbf{k}$-resolved rt-TDDFT algorithm is inherited from an earlier single-$\Gamma$ version of Time Dependent \textit{ab initio} Package (\texttt{TDAP}),~\cite{Meng2008} which is based on the \texttt{SIESTA}~\cite{Soler2002, Ordejon1996} package. In such a rt-TDDFT algorithm, the
    flowchart of a given ionic step is shown in Fig.~\ref{flowchart}. Each process is described in detail in the Secs. II A-II G, marked with the same labels as in Fig.~\ref{flowchart}. Here atomic units $\hbar = m_e = e = 1$ are used throughout this work.
    
    \subsection{Hamiltonian and overlap matrix}
    Adopting periodical boundary conditions, the lattice of an extended system are denoted as $\mathbf{R}_s$ ($s = 1, 2, 3, ..., N$) and the atoms $i$ in the unit cell are located at positions $\mathbf{b}_i$, where $N$ is truncated to construct a finite supercell. A set of numerical atomic-centered orbitals (NAOs) $\{\xi_{i\alpha}\}$ is associated with each atom in the simulated system, where $\alpha$ denotes both the orbital and angular quantum number of an atomic orbital,  each expressed in multiple radial basis functions $\zeta$~\cite{Soler2002}. Here, since all the operators and functions are time-dependent, we only denote the explicit dependence on $t$ as $f(t)$ and omit $t$ for implicit dependence.
    
    Overlap matrix $S_\mathbf{k}$ and Hamiltonian $H_\mathbf{k}$ at the each $\mathbf{k}$ point in the reciprocal space are expressed with NAOs:
    \begin{equation}
    \label{Sk}
    S_{i\alpha,j\beta,\mathbf{k}} = \sum_s e^{-i \mathbf{k} \cdot \mathbf{R}_s}
    \braket{\xi_{i\alpha}(\mathbf{r} + \mathbf{R}_s + \mathbf{b}_i) | \xi_{j\beta}(\mathbf{r} + \mathbf{b}_j)},
    \end{equation}
    \begin{equation}
    H_{i\alpha,j\beta,\mathbf{k}} = \sum_s e^{-i \mathbf{k} \cdot \mathbf{R}_s}\braket{\xi_{i\alpha}(\mathbf{r} + \mathbf{R}_s + \mathbf{b}_i) |\hat{H}| \xi_{j\beta}(\mathbf{r} + \mathbf{b}_j)},
    \end{equation}
    where 
    \begin{equation}
    \label{hamiltonian}
    \begin{split}
    	\hat{H} =& \hat{T} + \sum V_I^{local}(\mathbf{r}) + \sum V_I^{KB} + V^H(\mathbf{r}, \rho(\mathbf{r}))\\ 
    	&+ V^{XC}(\mathbf{r}, \rho(\mathbf{r})) + V^{ext}(\mathbf{r})
    \end{split}
    \end{equation}
    is the Hamiltonian {operator}. Here $\hat{T} = \frac{1}{2}\nabla_{\mathbf{r}}^2$ is the kinetic energy operator, $I$ is the index for atoms, $V_I^{local}$ and $V_I^{KB}$ are the local and Kleinman-Bylander parts of the pseduopotential for the $I$th atom, $V^H$ is the Hartree potential, $V^{XC}$ is the exchange-correlation (XC) potential and $ V^{ext}$ is the potential of external field. Details in the calculation of $\braket{\xi_{i\alpha}(\mathbf{r} + \mathbf{R}_s + \mathbf{b}_i)  |\hat{H}| \xi_{j\beta}(\mathbf{r} + \mathbf{b}_j)}$ are described in Ref.~\onlinecite{Ordejon1996}. Within adiabatic local density approximation (LDA) and generalized gradient approximations (GGA)~\cite{Yabana1996} for the exchange-correlation functional, $V^{XC}$ does not depend explicitly on time $t$, i.e. $V^{XC}[\rho(\mathbf{r},t),t] = V^{XC}[\rho(\mathbf{r},t)]$. Thus, most XC functionals in ground-state DFT such as Perdew-Wang~\cite{Perdew1981}, Perdew-Burke-Ernzerhof~\cite{Perdew1996}, Becke-Lee-Yang-Parr~\cite{Becke1988, Lee1988}, and van der Waals density functional~\cite{Dion2004, Roman-Perez2009} are compatible in this implementation of rt-TDDFT.
    
    \subsection{External field}
    To simulate the laser-matter interactions, time-dependent electric field $\mathbf{E}(t)$ is introduced to the Hamiltonian to represent the external time-dependent laser field in two different scenarios: the length gauge and vector gauge. 
    
    Within the length gauge, the effect of electric field $\mathbf{E}(t)$ is added to $V^{ext}$ as a scalar potential
    \begin{equation}
    V^{ext}(\mathbf{r}, t) = - \mathbf{E}(t)\cdot \mathbf{r}.
    \end{equation}
    Time dependent $\mathbf{E}(t)$ can be tuned adopting any shape in time evolution. A most popular example is using the shape of a Gaussian wave packet
    \begin{equation}
    \label{eq:GaussianWave}
    \mathbf{E}(t)=\mathbf{E}_0\cos\left(2\pi f t + \phi\right)\exp\left[-\frac{(t-t_0)^2}{2\sigma^2}\right],
    \end{equation}
    where $f$ is the laser frequency, $t_0$ is the peak time, and $\phi$ is the phase factor. 
    
    We note that, the translational symmetry of Hamiltonian is broken by the introduction of finite external field $\mathbf{E}$ in the length gauge, since 
    \begin{equation}
    V^{ext}(\mathbf{r} + \mathbf{R}_s, t) = - \mathbf{E}(t) \cdot (\mathbf{r} + \mathbf{R}_s) \neq - \mathbf{E}(t) \cdot \mathbf{r}.
    \end{equation}
    Thus, a common solution is using a sawtooth field along spatial direction ${\mu} \in \{x,y,z\}$ 
    \begin{equation}
    E_{\mu}(\mathbf{r},t)=
    \begin{cases}
    {E_{\mu}(t)}&  \epsilon < x_{\mu} < L_\mu - \epsilon,\\
    - {E_{\mu}(t)}L_{\mu}/2\epsilon &  -\epsilon < x_{\mu} < + \epsilon.
    \end{cases}
    \end{equation}
    where $L_\mu$ is the length of unit cell along ${\mu}$ and $\epsilon \rightarrow 0$. Thus, $-{E_{\mu}(t)}L_{\mu}/2\epsilon \rightarrow \infty$, which requires that charge density vanishes $\rho(x_\mu) = 0$ in the region $-\epsilon < x_{\mu} < + \epsilon$, otherwise the energy diverges. Thus, a vacuum layer is essential along ${\mu}$. {The requirement for a vacuum layer limits the application of theoretical approaches using the length gauge field to study the extended systems. Since there is no vacuum layer in the extended bulk systems, the translational symmetry of the Hamiltonian is broken, $H(\mathbf{r} + \mathbf{R}_s) \neq H(\mathbf{r})$, using the length gauge field.} {Plus, length gauge field is invalid in large systems and in short wavelength perturbation~\cite{Lestrange2015}.}
    
    Dynamical electric field in the vector gauge by introducting vector potential $\mathbf{A}$ could preserve the transitional symmetry of Hamiltonian, thus removes the requirement of the vacuum layer.~\cite{Yabana2006a, Yabana2012} The relation between $\mathbf{E}$ and $\mathbf{A}$ is
    \begin{equation}
    \mathbf{E} = -\frac{1}{c}\frac{\partial \mathbf{A}}{\partial t}; \mathbf{A} = -c\int \mathbf{E} dt.
    \end{equation}
    	The Hamiltonian with the presence of $\mathbf{A}$ is then
    	\begin{equation}
    	H = \frac{1}{2m} (\hbar\mathbf{k} - \frac{e}{c}\mathbf{A})^2 = \frac{1}{2m} (\hbar k + e \int \mathbf{E} dt)^2 = \frac{\hbar^2}{2m} ( \mathbf{k} + \mathbf{k_A} )^2,
    	\end{equation} 
    	where
    	\begin{equation}
    	\mathbf{k_A} = \frac{e}{\hbar} \int \mathbf{E} dt = \sqrt{2} \int \mathbf{E} dt.
    	\end{equation}
    	within Rydberg atomic unit, where $e = \sqrt{2}$, $\hbar = 1$ and $t = \hbar$/Ry.
    	The unit of $\mathbf{k_A}$ is $\mathrm{Bohr}^{-1}$, the same as the unit of $\mathbf{k}$.
    	
    	
    	\subsection{Propagation}
    	With time-dependent (TD) Hamiltonian and overlap matrix, TDKS equation is solved to obtain $\ket{u_{n\mathbf{k}}(\mathbf{r},t+\Delta t)}$ from the state $\ket{u_{n\mathbf{k}}(\mathbf{r},t)}$ at the previous time step: \begin{equation}
    	\label{evolution}
    	\ket{u_{n\mathbf{k}}(\mathbf{r}, t_2)} = \exp\left[-i S^{-1}_\mathbf{k}(t') H_\mathbf{k}(t')\Delta t \right] \ket{u_{n\mathbf{k}}(\mathbf{r}, t_1)}.
    	\end{equation} 
    	where $\Delta t = t_2 - t_1$ is the length of time step, $\ket{u_{n\mathbf{k}}(\mathbf{r},t)}$ is Bloch function and
    	$t' \approx ({t_1 + t_2})/{2}$. 
    	
    	It is rather difficult to evaluate $H_\mathbf{k}(t')$ and $S_\mathbf{k}(t')$ directly. Because $\Delta t$ is quite small ($< 0.05$~fs), the ion positions barely changes from $t_1$ to $t_2$. Since $S_\mathbf{k}(t)$ is only determined by ionic positions (Eq.~(\ref{Sk})), it is accurate enough to assume $S_\mathbf{k}(t') \approx S_\mathbf{k}(t_2)$. However, $H_\mathbf{k}(t)$ may largely change due to the rapid evolution of electrons. {To approximate $H_\mathbf{k}(t')$ properly, mid-point technique has been widely used~\cite{Li2005a,Goings2018}}.
    	
    	
    	{Note that, $\ket{u_{n\mathbf{k}}(\mathbf{r})(t_2)}$ is not explicitly dependent on other TDKS orbitals $\ket{u_{n'\mathbf{k}'}(\mathbf{r})(t_1)}$ ($n' \neq n$ or $\mathbf{k}' \neq \mathbf{k}$), as a result of the $v$-representativity of the TDKS equations~\cite{Runge1984, Marques2012}. It decouples the evolution equations of different TDKS orbitals and make TDDFT calculations practical. However, it  nevertheless can account for both interband and intraband scatterings. Because $H_\mathbf{k}$ is determined by the total charge density, which is a weighted summation of all the occupied orbitals, there still exists an implicit coupling between different TDKS orbitals.}
    	
    	Numerically, the time propagator $\exp(-i S^{-1}_\mathbf{k}H_\mathbf{k}\Delta t)$ in Eq.~(\ref{evolution}) is expanded using first-order Crank-Nicholson scheme:
    	\begin{equation}
    	\label{Crank-Nicholson}
    	\ket{u_{n\mathbf{k}}(\mathbf{r}, t_2)} = \frac{1-i S^{-1}_\mathbf{k}H_{\mathbf{k}} \Delta t/2}{1+i S^{-1}_\mathbf{k}H_{\mathbf{k}} \Delta t/2} \ket{u_{n\mathbf{k}}(\mathbf{r}, t_1)}.
    	\end{equation}
    	Technically, since computing $S^{-1}_\mathbf{k}$ is the most time-consuming part in the calculation of Eq.~(\ref{Crank-Nicholson}), we minimize the times for its computing. $S^{-1}_\mathbf{k}$ is only updated when atomic positions, thus the center of NAOs, $\mathbf{b}_i$ are changed. Consequently, when ions are fixed, $S^{-1}_\mathbf{k}$ is computed only once at the first ionic step. Even with ions moving, $S^{-1}_\mathbf{k}$ only need to be updated once for each ionic step.
    	
    	\subsection{Updating charge density}
    	With $\ket{u_{n\mathbf{k}}(\mathbf{r},t_2)}$ solved in Eq.~(\ref{evolution}), 
    	the density matrix (DM) $\rho_{i\alpha,j\beta}(t_2)$ is computed accordingly as:
    	\begin{equation}
    	\begin{split}
    	\rho_{i\alpha,j\beta}(t_2) &= \sum_n\sum_\mathbf{k} q_{n,\mathbf{k}} \ket{u_{n\mathbf{k}}(\mathbf{r}, t_2)}\bra{u_{n\mathbf{k}}(\mathbf{r}, t_2)}\\
    	& = \sum_n\sum_\mathbf{k} q_{n,\mathbf{k}} c^*_{n,i\alpha,\mathbf{k}}(t_2) c_{n,j\beta,\mathbf{k}}(t_2),
    	\end{split}
    	\end{equation}
    	where $q_{n,\mathbf{k}}$ is the electronic population of the band $n$ at $\mathbf{k}$, and $c_{n,j\beta,\mathbf{k}}(t_2)$ is the coefficient of 	$\ket{u_{n\mathbf{k}}(\mathbf{r}, t_2)}$ in NAO basis:
    	\begin{equation}
    	\ket{u_{n\mathbf{k}}(\mathbf{r}, t_2)} = \sum_{j\beta} c_{n,j\beta,\mathbf{k}}(t_2) \xi_{j\beta}(\mathbf{r}).
    	\end{equation}
    	
    	\subsection{Self-consistent evolution}
    	We use the self-consistent process described in Ref.~\cite{Meng2008} during the time evolution of charge density. This process substantially increases the numerical stability~\cite{Ren2010a}. All the criteria for convergence test developed in SIESTA are compatible with the current approach, such as using the maximum element of the DM difference, the energy difference, or the Harris energy difference, etc. as a criterion for achieving self-consistency.~\cite{Soler2002}.
    	
    	Here, we use DM difference as an example. Convergence in charge density during time evolution is reached when
    	\begin{equation}
    	\max\left\{\left|\rho^{new}_{i\alpha,j\beta} - \rho_{i\alpha,j\beta}\right|\right\} < \eta,
    	\end{equation}
    	where $\eta$ is about 10$^{-4}$. 
    	
    	\subsection{Mixing}
    	If not converged, the linear mixing of DM is needed to generate the input DM for computing charge density $\rho_{next}$ at the next cycle, instead of using $\rho_{new}$ directly,
    	\begin{equation}
    	\label{eq-mix}
    	\rho = (1-w)\rho + w\rho_{new},
    	\end{equation}
    	where the $\rho$ on the right side of Eq. (\ref{eq-mix}) is the input DM and $\rho_{new}$ is the output DM, and $w$ is the mixing weight, usually $w = 0.1 - 0.5$.
    	
    	\subsection{Postprocessing}
    	If self-consistent time evolution of charge density is converged, the postprocessing steps are evoked, including the calculation of total energy, Hellmann-Feynman forces, ionic movements, etc. These functions are implemented in SIESTA~\cite{Soler2002} and compatibly used in TDAP~\cite{Meng2008}. {We note that, rt-TDDFT in atomic orbital basis gives rise to additional Pulay terms that contribute to the force evaluations\cite{Ding2015a,Isborn2007a,Schlegel2001}. The total force is the combination of Hellmann-Feynman force and Pulay term.} {With the calculated forces, the coupled electron-ion motion can be simulated based on classical ionic trajectories, in the framework of Ehrenfest dynamics. In Ehrenfest dynamics, the forces on the ions are averaged over the adiabatic electronic states along all possible ionic paths. If one path is dominating or many similar potential energy surfaces are involved, Ehrenfest dynamics works very well~\cite{Tully1990}; otherwise, classical trajectory approximations in Ehrenfest dynamics become less accurate~\cite{Parandekar2006, Hack2000}. Furthermore, detailed balance for quantum electronic states is absent in the Ehrenfest dynamics. Thus, the applications of the present method are limited to the  cases where the averaged potential energy surfaces yields a reasonable description of coupled electron-ion dynamics. Since we focus on the dynamics of excited electrons in this work, the ions are fixed in the simulations.}
    	
        {Here we introduce some analysis in detail for typical rt-TDDFT simulations. First, we could evaluate the state-to-state transition probabilities between TDKS orbitals during time evolution~\cite{Li2005a, Rohringer2006a}}:
    	\begin{equation}
    	\label{projection}
    	P_{nn'\mathbf{k}} = \left|C_{nn'\mathbf{k}} \right|^2 = \left|\braket{v_{n\mathbf{k}}|S_k|u_{n'\mathbf{k}}}\right|^2,
    	\end{equation}
    	where $\ket{v_{n\mathbf{k}}}$ is the adiabatic basis satisfying
    	\begin{equation}
    	\label{adiabaticBasis}
    	H_\mathbf{k}\ket{v_{n\mathbf{k}}(\mathbf{r})} = E_{n\mathbf{k}} S_\mathbf{k}\ket{v_{n\mathbf{k}}(\mathbf{r})}.
    	\end{equation}
    	The population $\mathcal{q}_{n\mathbf{k}}$ of the adiabatic state $n\mathbf{k}$ is thus projected from the TDKS orbitals at a given time as:
    	\begin{equation}
    	\label{eq:excitation}
    	\mathcal{q}_{n\mathbf{k}} =  \sum_{n'\in n_{\mathbf{k},occ}} q_{n'\mathbf{k}} P_{nn'\mathbf{k}},
    	\end{equation}
    	where $n_{\mathbf{k},occ}$ is the occupied state at $\mathbf{k}$ point.
    	

    	For finite systems and surface slabs, we can calculate time-dependent dipole moment along the direction.
    	For periodic systems, the dipole moment is ill-defined. Instead, we calculate time dependent current,   
    	\begin{equation}
    	\mathbf{j} = -i\frac{e\hbar}{m} \sum_n ( \braket{u_{n\mathbf{k}}|\nabla|u_{n\mathbf{k}}} - \braket{u_{n\mathbf{k}}|\nabla|u_{n\mathbf{k}}}^*  ),
    	\end{equation}
    	as the response function.
    	
    	
    	
    	\section{Results and Discussion}
    	\subsection{Momentum-resolved versus supercell approaches}
    	\begin{figure}
    		\centering
    		\includegraphics[width = \fitwidth]{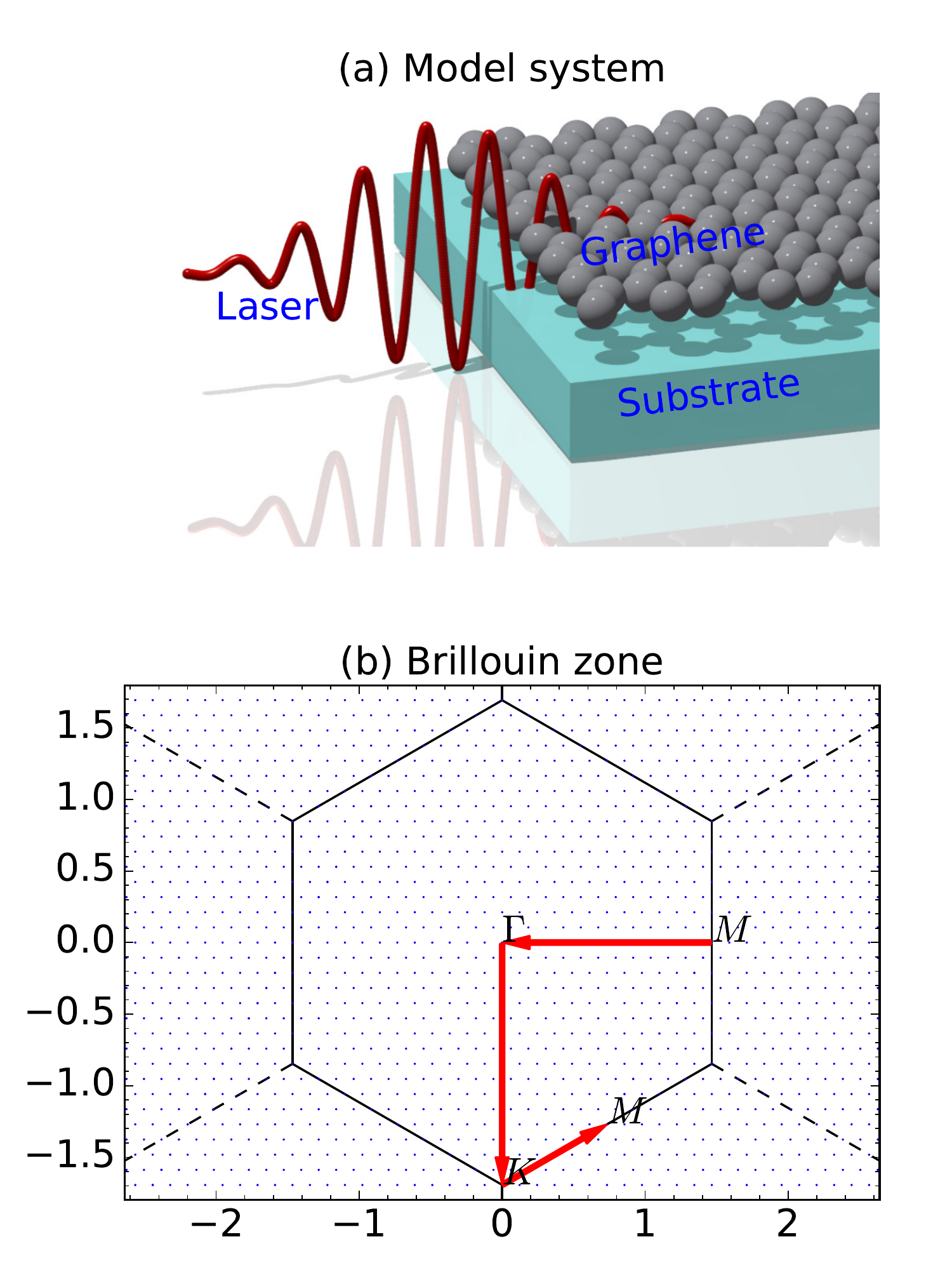}
    		\caption{(a) Sketch of graphene under out-of-plane polarized laser field. (b) Brillouin zone and $\mathbf{k}$ sampling of graphene. Blue dots denote the $\mathbf{k}$ points used for sampling. Red lines denote the high symmetric path.}
    		\label{struct}
    	\end{figure}
    	To demonstrate the $\mathbf{k}$-resolved algorithm, we choose graphene as the model system (see Fig.~\ref{struct}(a)). An exotic property of graphene (also of other Dirac materials) is the linear dispersion near K point, namely, $E(\mathbf{k}) = v_F\mathbf{k}$, where $E$ the band energy and $v_F$ is the Fermi velocity which could reach $10^6$~m/s. To describe all the Bloch electrons, especially those near the Fermi energy, two kinds of strategies are used: unit cell calculations with $\mathbf{k}$-resolved reciprocal space samplings or a supercell approach with $\Gamma$-only $\mathbf{k}$-sampling. To demonstrate the advantages of the $\mathbf{k}$-resolved algorithm, we compare three cases: 
    	\\ (i) unit cell with the Monkhorst-Pack~\cite{monkhorst1976special} $N_k \times N_k \times 1$ $\mathbf{k}$ point mesh, to cover all important special $\mathbf{k}$-points $M$, $\Gamma$ and $K$, facilitating a line-mode analysis along $M \rightarrow \Gamma \rightarrow K \rightarrow M$ [Fig.~\ref{struct}(b)]; 
    	\\ (ii) $N_c\times N_c \times 1$ supercell with single $\Gamma$ point; and \\
    	(iii) $N_c\times N_c \times 1$ supercell with single $K$ point.
    	
    	To compare the computation accuracy of these three cases, we define an error function $\Delta$ as,
    	\begin{equation}
    	\label{delta}
    	\Delta = \frac{1}{T}\int_0^T {|E_{ex}(t) - E^{ref}_{ex}(t)|}dt,
    	\end{equation}
    	where $T$ is the total simulation time, $E^{ref}_{ex}$ is the excitation energy of the reference case and $E_{ex}(t)$ is the excitation energy
    	\begin{equation}
    	\label{eq:ExcitationEnergy}
    	E_{ex}(t) = E_{KS}(t) - E_{KS}(t=0),
    	\end{equation}
    	where $E_{KS}$ is the total energy of the system. 
    	\begin{figure}
    		\centering
    		\includegraphics[width=\fitwidth]{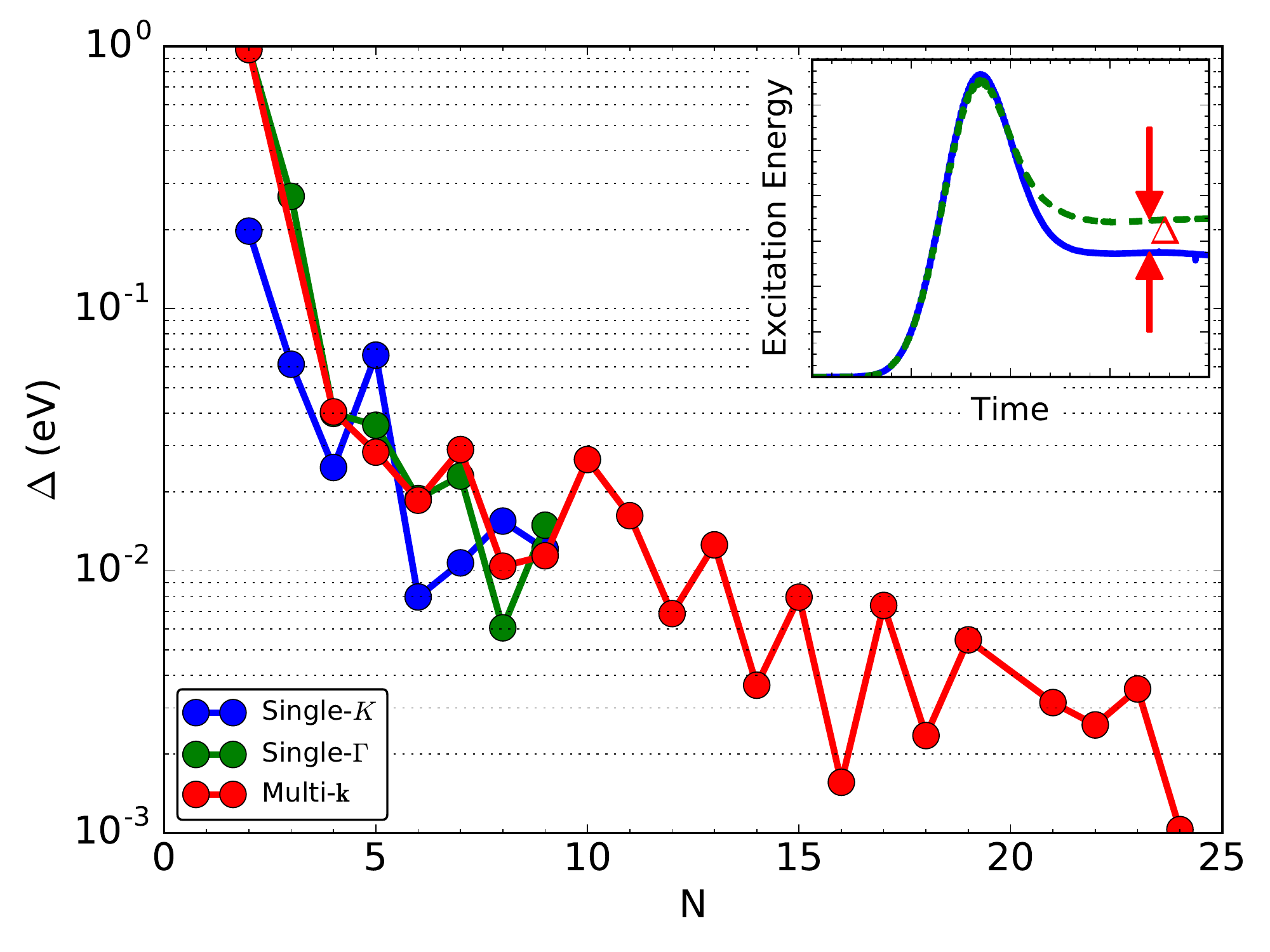}
    		\caption{Error function $\Delta$, as defined in Eq.~(\ref{delta}), as a function of N with different supercell and $\mathbf{k}$ mesh set-ups. $N$ denotes $N_k$ for $\mathbf{k}$-resolved approach, or $N_c$ for the single K and single $\Gamma$ supercell approach.}
    		\label{fig:plotDeltaGauge}
    	\end{figure}
    	
    	Here, we evaluate $\Delta$ under such settings: the Gaussian-shaped laser pulse [Eq. (\ref{eq:GaussianWave})] with $\phi = 0$, $t_0 = 7.0$~fs, $\sigma = 2.0$~fs, $f = 21.93$~eV is applied; the total simulation time is $T = 20$~fs; and the reference energy $E^{ref}_{ex}$ is calculated with $60\times60\times1$ $\mathbf{k}$-point mesh. A diagram to illustrate the definition of $\Delta$ is shown in the inset of Fig.~\ref{fig:plotDeltaGauge}. {The time step is chosen to $\Delta t = 0.02$~fs and the total time is $20$~fs. Troullier-Martin pseudopotentials~\cite{Troullier1991}, adiabatic local density approximation (ALDA) exchange-correlation functional~\cite{Perdew1981, Yabana1996} and an auxiliary real-space grid equivalent to a plane-wave cutoff of $75$~Ry are used. In description of C atoms, we use a basis set of 8 double-$\zeta$ orbitals \{2s(2$\zeta$), 2p$_x$(2$\zeta$), 2p$_y$(2$\zeta$), 2p$_z$(2$\zeta$)\} and 5 polarization orbitals \{ $\mathrm{P_{d_{xy}}}$, $\mathrm{P_{d_{yz}}}$, $\mathrm{P_{d_{z^2}}}$, $\mathrm{P_{d_{xz}}}$, $\mathrm{P_{d_{x^2-y^2}}}$ \}. We calculate the test cases with one 8-core Intel(R) Xeon(R) CPU E5-2650@2.00GHz.} 
    	
    	We plot $\Delta$ of these three cases in Fig.~\ref{fig:plotDeltaGauge}. The error $\Delta$ decreases as $N = N_k$ (or $N_c$) increases. The absolute value of $\Delta$ on the the same scale is achieved with $N_c = N_k$. That is to say, the unit cell approach with $N \times N \times 1$ $\mathbf{k}$-point mesh is as accurate as the approach using a $N\times N \times 1$ supercell. To achieve an accuracy with the $\Delta \le 2$~meV/atom, $N_k  = 24$ is needed. Thus, it can be predicted that $N_C = 24$ is needed for the supercell approach. 
    	
    	\begin{figure*}
    		\centering
    		\includegraphics[width=\linewidth]{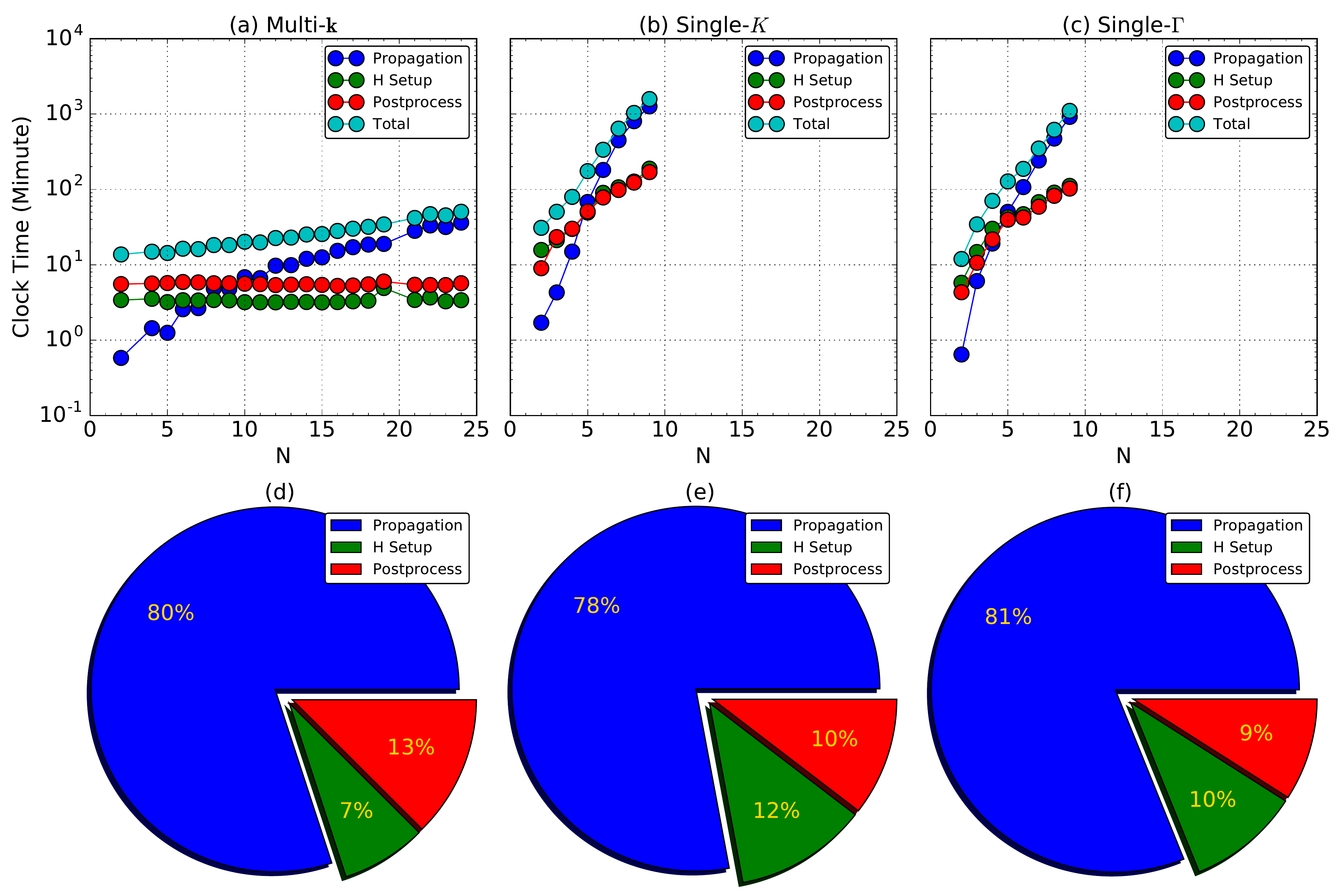}
    		\caption{(a) -- (c) CPU clock time as a function of $N_k$ or $N_c$. \textit{H setup} corresponds to the calculation in Eq.~(\ref{hamiltonian}), \textit{Postprocesses} mainly comprises calculating Hellmann-Feynman forces, and \textit{Propagation} corresponds to computation in Eq.~(\ref{evolution}). (d--f) Percentages of the computer time consumption for each process with (d)$N_k = 24$, and (e-f) $N_c = 9$. }
    		\label{fig:noProj-K-plotTimeScan}
    	\end{figure*}
    	However, we emphasize that the computational cost for calculating $N_c\times N_c \times 1$ supercell is extremely heavy. As shown in Fig.~\ref{fig:noProj-K-plotTimeScan}, solving Eq.~(\ref{evolution}) dominates ($\sim 80$\%) the computer time consumption at large $N_k$ ($N_c$), which scales linearly with the total number of $\mathbf{k}$ points, $N_k^2$, and quadratically with the total number of atoms, $N_c^2$. The CPU clock time $t_{c}$ approximately scales as O($N^2_k\times N_c^4$). Thus, $t_c = N_c^4$ for supercell calculation, while $t_c = N_k^2$ for$\mathbf{k}$-resolved calculations at the same level of accuracy. 
    	
    	As $N$ increases, this difference become more significant. For supercell calculations, we are able to only compute supercells up to $N_C = 9$, which already costs over $2\times10^3$~min. At the same accuracy level, $N_k = 9$ calculation costs only $20$~min, which is only 1/100 of that for $N_C = 9$ case, consistent with the time complexity analysis $N_k^2/N_c^4 = 1/81$. As mentioned above, $N_k  = 24$ or $N_C = 24$ is needed for relatively accurate calculations. To fulfill this requirement, calculation with $N_k = 24$ costs only about 1 hour, showing that it is readily accessible and efficient. In contrast, calculating a $N_c = 24$ supercell would require a computer time over 576 hours (24 days) and thus heavy in real applications. Regarding the computational accuracy and efficiency, we choose $\mathbf{k}$-point mesh $24\times24\times1$ to achieve an extremely dense sampling of the Brillouin zone. 
    	
    	{With the small unit cell of graphene, the number of real space grids is $\sim$1000, which is 30 times of the number of NAOs used. Considering the evolution algorithm has the computational complexity of $O(n^2)$, where $n$ is the number of basis functions, the computer time for wavefunction evolution using NAO basis is largely reduced to 1/90 of that using real space grid basis. In practical calculations using the same number of message-passing-interface (MPI) processes, the reduction in the total computer time is tested to be about 1/5 to 1/10 depending on the systems under consideration~\cite{Meng2008,Lian2018}.}
    	
    	
    	\subsection{Out-of-plane excitation}
    	We then adopt a laser field perpendicularly polarized to the graphene plane to excite electrons in graphene, i.e. in a set-up of small angle scattering. Since there is a vaccum layer along the out-of-plane direction, the laser field in the length gauge can be used. 
    	
    	We first calculate the dielectric function of graphene to locate the photon energy for resonant excitation, $\alpha_{\mu,\nu}$, where $\mu$, $\nu$ denote the spatial direction $\mu$, $\nu$ $\in$ \{$x$, $y$, $z$ \}. The $\alpha_{\mu,\nu}$ describes the response of dipole moment $P_\mu(\omega)$ to the electric field $E_\nu(\omega)$ in the frequency domain,
    	\begin{equation}
    	\label{eq:response}
    	P_\mu(\omega) = \alpha_{\mu,\nu}(\omega) E_\nu(\omega).
    	\end{equation}

    	In rt-TDDFT calculations, we apply the electric field $E_\nu(t)$ and obtain the dipole moment $P_\mu(t)$ in time domain. Then we carry out the Fourier transform to get Eq.~[\ref{eq:response}],
    	\begin{equation}
    	\int P_\mu(t)\exp(i\omega t) dt = \alpha_{\mu,\nu}(\omega) \int  E_\nu(t)\exp(i\omega t) dt.
    	\end{equation}
    	We then obtain
    	\begin{equation}
    	\alpha_{\mu,\nu}(\omega) = \frac{\int  P_\mu(t)\exp(i\omega t) dt}{\int  E_\nu(t)\exp(i\omega t) dt}.
    	\end{equation}
    	In principle $E_\nu(t)$ can be in an arbitrary shape with time. However, in practice, it is better to choose Dirac function $E^\delta_\nu(t) = E_{\nu0} \delta(t)$, or the Heaviside step function $E^\theta_\nu(t) = E_{\nu0} [1 - \theta(t)]$ to include components $E_\nu(\omega)$ at all $\omega$, since we have
    	\begin{eqnarray}
    	E^\theta_{\nu}(\omega) = \int E_{\nu0}[1 - \theta(t)]\exp(i\omega t) dt = \frac{E^0_{\nu}}{i\omega}.
    	\end{eqnarray}
    	Here we choose the latter form:
    	\begin{equation}
    	\label{stepEField}
    	E^\theta_\nu(t)= E^0_\nu [1 - \theta(t)] = \begin{cases} E^0_\nu& t \le 0  \\ 0& t > 0 \end{cases},
    	\end{equation}
    	which leads to 
    	\begin{equation}
    	\label{eq:imagDielecFunc}
    	\alpha_{\mu,\nu}(\omega) = \frac{i\omega}{E^0_{\nu}}\int P_\mu(t)\exp(i\omega t) dt.
    	\end{equation}
    	Importantly, $\mathrm{Im}\{\alpha_{\mu,\mu}(\omega)\}$ characterizes the optical absorbance at $\omega$ along the $\mu$ direction. 
    	\begin{figure}
    		\centering
    		\includegraphics[width=\fitwidth]{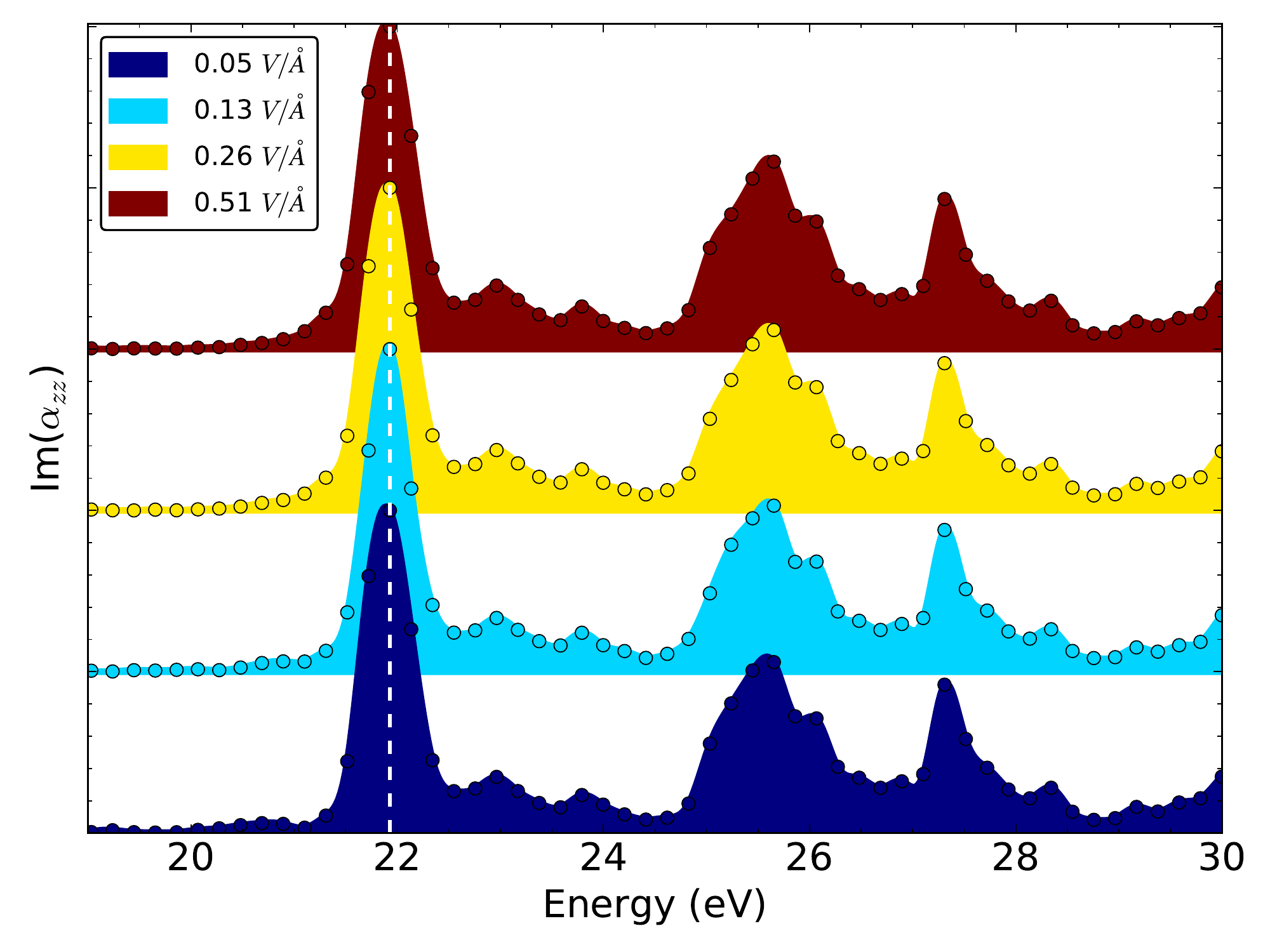}
    		\caption{The $\mathrm{Im}\{\alpha_{\mu,\mu}(\omega)\}$ as a function of photon energy $\omega$ at different $E^0_z$ using Eq.~(\ref{eq:imagDielecFunc}). }
    		\label{fig:dielectriFunction}
    	\end{figure}
    	
    	We calculate the imaginary part of the dielectric function along the out-of-plane $z$ direction of graphene,  $\mathrm{Im}\{\alpha_{z,z}(\omega)\}$. As shown in Fig.~\ref{fig:dielectriFunction}, $\mathrm{Im}\{\alpha_{z,z}(\omega)\}$ are almost the same with the increase of $E^0$ from $0.05$ to $0.5$~V/\AA, indicating the linear response theory is appropriate in this range of light illumination. The absorption peaks are located at relative high energies ($> 20$~eV). The first absorption peak is located at $21.93$~eV. We choose this photon energy to simulate the resonant excitation of graphene in the perpendicular direction. 
    	
    	\begin{figure}
    		\centering
    		\includegraphics[width = \fitwidth]{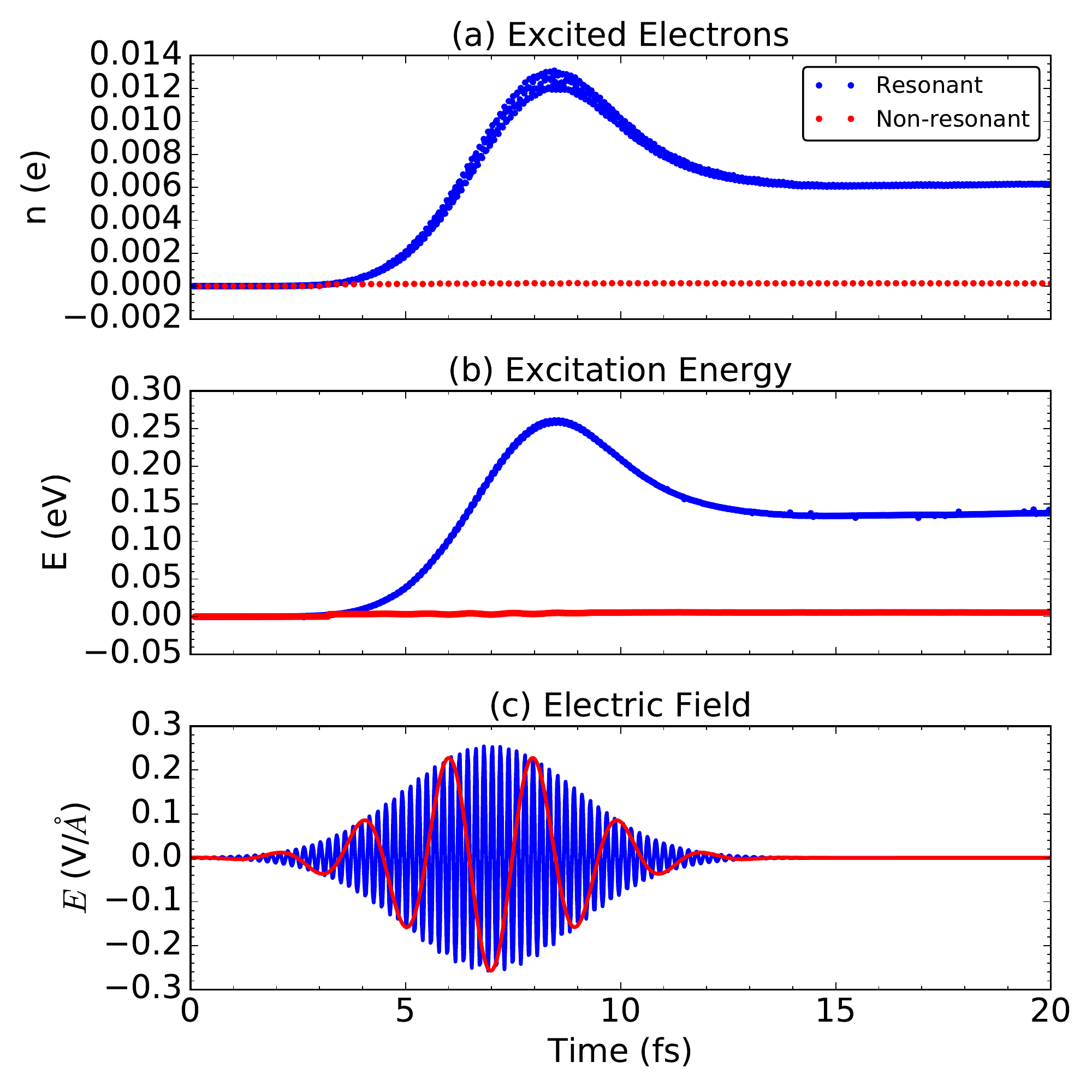}
    		\caption{(a) The number of excited electrons, (b) the total energy during excitation, and (c) the profile of laser field as a function of time. Red curves are for the non-resonant case at the photon energy $\omega_{nr} = 2.0$~eV, while blue curves are for the resonant case at $\omega_r = 21.93$~eV.}
    		\label{ExcitationVsTimeGaussian}
    	\end{figure}
    	We demonstrate the excitation dynamics of graphene at a resonant light frequency of $\omega_r = 21.93$~eV, and compare to the case at the non-resonant light frequency of $\omega_{nr} = 2.0$~eV. 
    	We characterize the overall excitation through tracking the number of excited electrons, as well as the total energy change during the excitation process as a function of time. The number of excited electrons $n(t)$ is calculated as,
    	\begin{equation}
    	\label{eq:ExcitedElectrons}
    	n(t) = \sum_{unocc} \mathcal{q}_{n\mathbf{k}}(t),
    	\end{equation}
    	where $\mathcal{q}_{n\mathbf{k}}(t)$ is obtained from Eq.~(\ref{eq:excitation}), and ${unocc}$ denotes the unoccupied TDKS states. 
    	
    	As shown in Fig.~\ref{ExcitationVsTimeGaussian}, different behaviors are observed for the two excitation conditions. The excited electrons $n(t)$ and excitation energy $E_{ex}(t)$ increases at $\omega_r$, while no response is observed at $\omega_{nr}$. The same results are obtained at other non-resonant light frequencies of $1.0$, $2.0$, $4.0$~eV. {It verifies that the calculated $\mathrm{Im}\{\alpha_{z,z}(\omega)\}$ characterizes well selectivity in optical absorption: only the light with the right photon energy $\omega$, at which  $\mathrm{Im}\{\alpha_{z,z}(\omega)\}$ peaks, has a strong absorption.}
    	
    	We discuss the resonant case here. In general, $n(t)$ is similar to the shape of the laser pulse, while two special features are observed. Firstly, the time variation in $n(t)$ has a $1.4$~fs delay from the laser field. This delay represents the intrinsic response time of graphene to laser field, namely, the time needed for light absorption and electronic transitions. {Secondly, $n(t)$ decreases but does not vanish after the end of light pulse.} Thus, we propose that two kinds of excitation process exist: one is the transient excited electrons, which quickly vanishes after the laser pulse is off; another is the residual excited electrons, which live relatively longer. 
    	Residual $n(t)$ would decrease with the occurrence of electron-electron and further electron-phonon scatterings at the time scale of $100$~fs, thus is not observed in our short-time simulation ($<20$~fs).
    	{We note that, the dependence on history is absent in the calculations with adiabatic exchange-correlation functionals, which causes less accurate prediction of the lifetime of excited states and ionic forces on a long time scale~\cite{PhysRevLett.89.023002, PhysRevLett.109.266404, doi:10.1002/qua.20465, doi:10.1063/1.2406069, doi:10.1063/1.4908133}.}
    	

    	\begin{figure}
    		\centering
    		\includegraphics[width = \fitwidth]{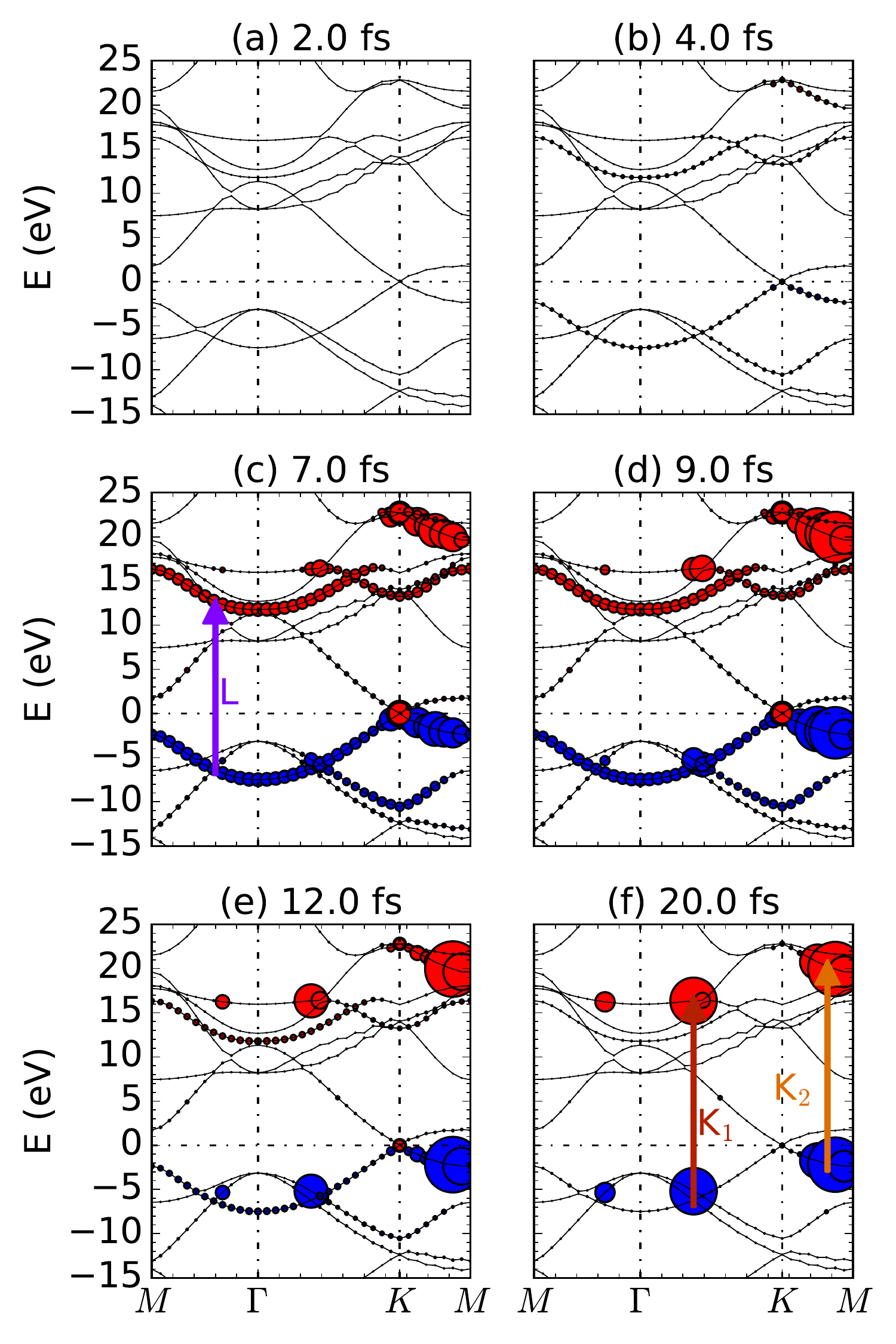}
    		\caption{Snapshots of excitation population $\Delta \mathcal{q}_{n\mathbf{k}}(t)$ at different time $t$. Black curves denote the time-dependent band structure of graphene. Blue cycles represent $\Delta \mathcal{q}_{n\mathbf{k}}(t) < 0$ and red cycles represent $\Delta \mathcal{q}_{n\mathbf{k}}(t) > 0$. Radius of circles are proportional to the value of $|\Delta \mathcal{q}_{n\mathbf{k}}(t)|$.}
    		\label{fig:projectionGaussian}
    	\end{figure}
    	
    	To verify our assumption about the existence of two kinds of excitation processes, we further distinguish the excitation with $\mathbf{k}$-point resolution. We choose six snapshots of $\mathcal{q}_{n\mathbf{k}}(t)$ defined in Eq.~(\ref{eq:excitation}), as shown in Fig.~\ref{fig:projectionGaussian}. At $t = 2.0$~fs with the absence of laser pulse, no excitation is observed at all $\mathbf{k}$ points. At $t = 4.0$~fs, the excitation is still ignorable, while the laser field is just turned on, due to the delay in electronic response we discussed above. At the peak time of the laser pulse  $t = 7.0$~fs, $\mathcal{q}_{n\mathbf{k}}(t)$ shows a significant distribution over many $\mathbf{k}$-points. We mark the dominant excitation mode as L, which involving bonding $\pi$ and antibonding $\pi$ bands. With $t$ increases from $7$~fs to $12$~fs, the L mode excitation rapidly decreases. In contrast, two new modes (labeled by their locations in the reciprocal space, K$_1$ and K$_2$) increases and become dominant. K$_1$ and K$_2$ modes maintain within $20$~fs while L mode gradually vanishes. Thus, with the assistance of newly developed $\mathbf{k}$-resolved algorithm, we are able to successfully distinguish these two kinds of excitation processes: L mode produce the transient excited electrons while K$_1$ and K$_2$ modes produce the residual excited electrons. 
    	
    	\begin{figure}
    		\centering
    		\includegraphics[width = \fitwidth]{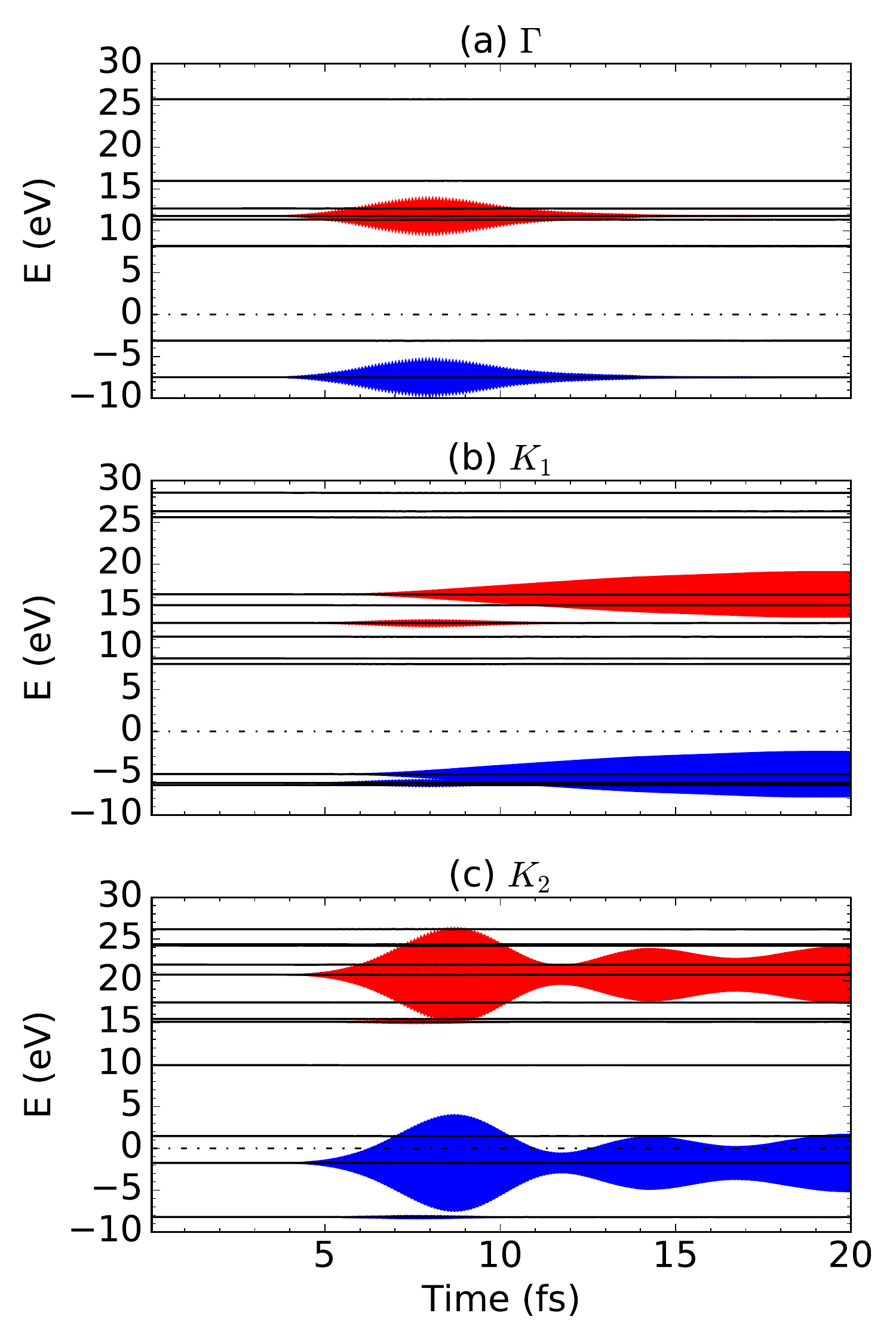}
    		\caption{Excitation population $\Delta \mathcal{q}_{n\mathbf{k}}(t)$ at different $\mathbf{k}$ as a function of time. Blue lines represent $\Delta \mathcal{q}_{n\mathbf{k}}(t) < 0$, and red lines represent $\Delta \mathcal{q}_{n\mathbf{k}}(t) > 0$ in population. Widths of the lines are proportional to $|\Delta \mathcal{q}_{n\mathbf{k}}(t)|$.}
    		\label{fig:populationSelectiveKvsTime}
    	\end{figure}
    	
    	Although K$_1$ and K$_2$ are both long-lived excitations, their time-dependence is quite different. We plot $\mathcal{q}_{n\mathbf{k}}(t)$ as a function of $t$ at three $\mathbf{k}$ points $\Gamma$, K$_1$ and K$_2$, as shown in Fig.~\ref{fig:populationSelectiveKvsTime}. For L mode (represented by photoexcitation at $\Gamma$ point), the clear transient character is demonstrated. The excitation only exists when the laser field is present, consistent with the observations in Fig.~\ref{fig:projectionGaussian}. However, for K$_1$ and K$_2$ modes, new differences are observed. Excited electrons in K$_1$ mode increases monotonically, while $\mathcal{q}_{n\mathbf{k}}(t)$ at K$_2$ shows an oscillation with a periodicity of $T_{K_2} \sim 5$~fs. These different behaviors are due to different excitation energies of three modes, originated from different band structures at the different $\mathbf{k}$-point. For instance, the oscillation of $K_2$ mode is analogous to the beating,
    	\begin{equation}
    	\mathcal{q}_{n{K_1}}(t) = A_0 \cos\left(\frac{\omega_{K_2} - \omega_r }{2} t\right) \cos\left(\frac{\omega_{K_2} + \omega_r }{2} t\right).
    	\end{equation}
    	For K$_2$ mode, $\omega_{K_2} = 22.75$~eV is the energy difference between the two electronic bands involved in the optical transition at K$_2$ (initial and final states), and $\omega_2 = \omega_{r} = 21.93$~eV is the driving photon energy. Beat frequency $T_b = {4\pi}/{(\omega_1 - \omega_2 )} = 5.07$~fs, which is close to the observed oscillation periodicity $T_{K_2}$. Thus, the oscillation of K$_2$ mode is the beat formed by the intrinsic band energy difference and the driving laser frequency. In contrast, $K_1$ mode excitation has very close energies: $\omega_{K_1} = 21.59$~eV and $ \omega_{r} = 21.93$~eV, thus only a half period of the beat ($T_{K_1} = 12.4$~fs) is observed in our simulation. For L mode, the excitation energy is $19.32$~eV, far below the $\omega_{r}$. A non-resonant interference shows up instead of beating. The rich photoexcitation phenomena discussed above and the associated complex dynamic behaviors hint for the needs for developing efficient rt-TDDFT algorithms with momentum resolution. By introducing a new degree of freedom in the reciprocal space, the $\mathbf{k}$-resolved dynamics labels the distinct excitation processes as well as final distribution of excited states in the Brillouin zone after the incidence of laser pulses.  
    	
    	\subsection{In-plane excitation}
    	\begin{figure}
    		\centering
    		\includegraphics[width=1.0\linewidth]{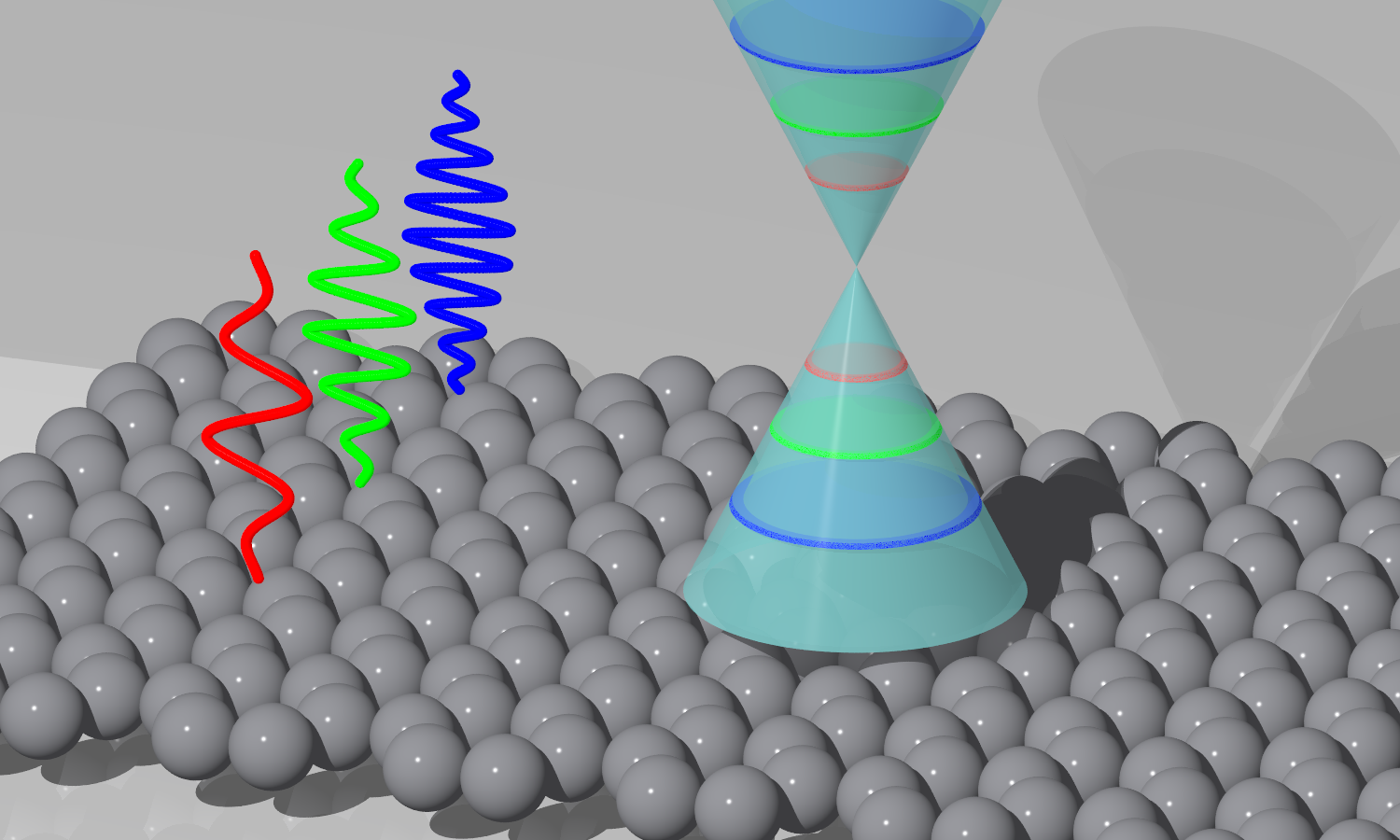}
    		\caption{Schematic of graphene excited by in-plane polarized laser pulse. The rings with different colors correspond to the electronic states involved in optical transitions introduced by the laser pulse with a different photon energy.}
    		\label{fig:model}
    	\end{figure}

    	For a laser pulse with its field polarization lying parallel to the atomic plane of graphene (i.e., normal incidence), adding a vacuum layer along the laser polarization direction is not possible for the periodical extended system such as graphene. Therefore we adopt the vector potential approach to simulate the in-plane laser-graphene interaction. The graphene sheet is illuminated with a linearly polarized laser pulse, as shown in Fig.~\ref{fig:model}. We note that in-plane excitation is well described by the Fermi's golden rule. Only the bands with an energy gap $\Delta E_g(\mathbf{k})$ equal to the photon energy $\omega$ will be excited. As a result, in-plane polarized laser excites electrons near the Dirac point for photon energies $\le$5 eV, see Fig.~\ref{fig:model}. The momentum-resolved simulation will distinguish the photoexcitation induced by a laser pulse with different photon energies $\omega$.
    	
\begin{figure}
    		\centering
    		\includegraphics[width=1.0\linewidth]{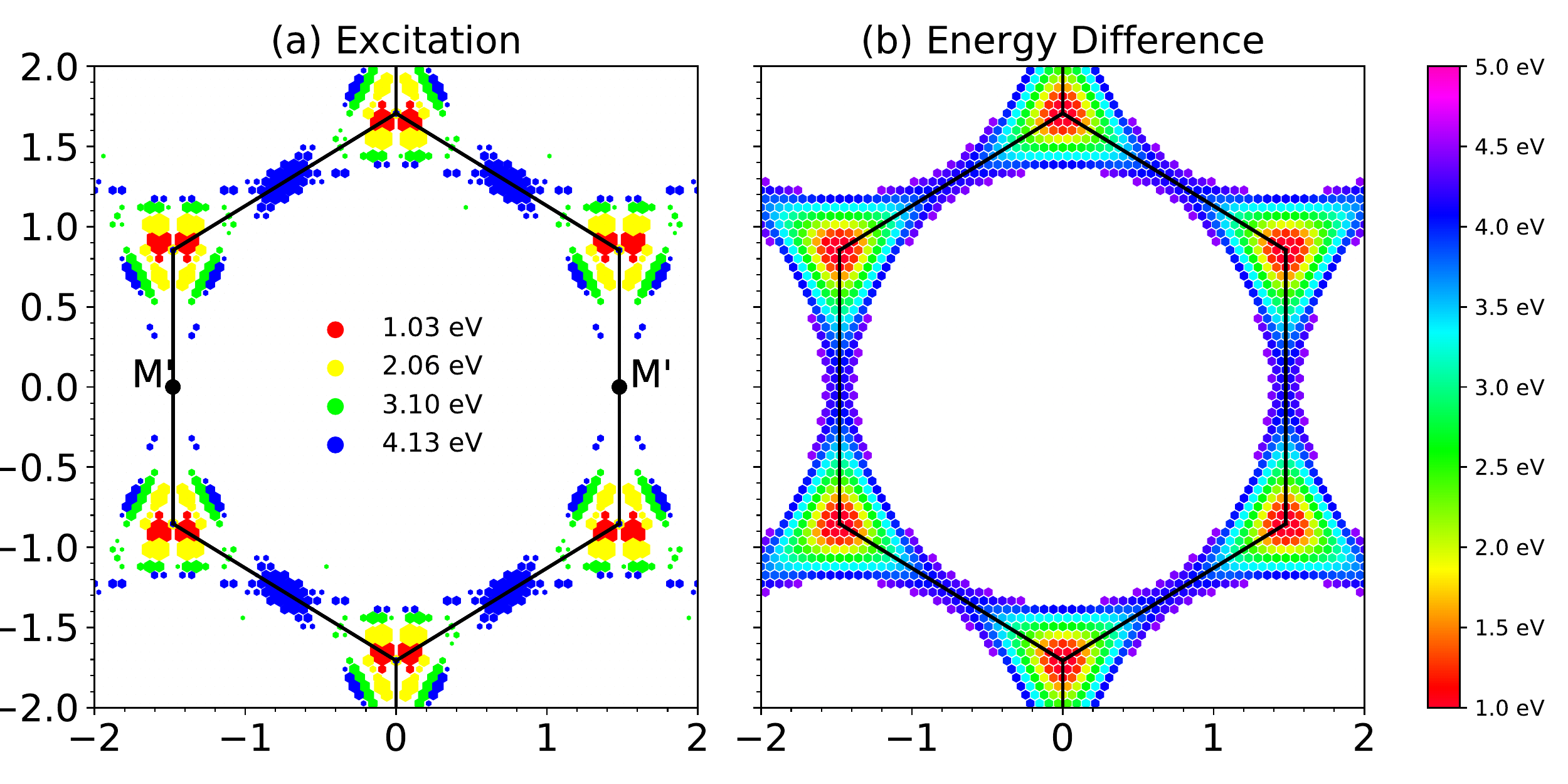}
    		\caption{Comparison of (a) distribution of excited electrons on different $\mathbf{k}$ points in the Brillouin zone of graphene,  and (b) corresponding energy differences in the electronic bands around the Dirac points.  }
    		\label{fig:nestingGrapheneAField}
    	\end{figure}

    	Here, we use four different wavelengths $\lambda =$ 1200 nm, 600 nm, 400 nm, and 300~nm for the laser pulse, corresponding to photon energies $\omega =$ 1.03, 2.06, 3.10, 4.13~eV, respectively, to excite graphene in the in-plane direction. For simplicity the laser field is polarized perpendicular to the C-C bond of graphene lattice (referred to as $y$ direction). The momentum resolved excitation patterns in the reciprocal space are shown in Fig.~\ref{fig:nestingGrapheneAField} (a), with the corresponding band energy difference $\Delta E_g(\mathbf{k})$ shown in Fig.~\ref{fig:nestingGrapheneAField} (b). It is clear that only the $\mathbf{k}$ points with $\Delta E_g(\mathbf{k}) = \omega$ are excited. This agreement justifies the validation of the vector gauge used in the current TDDFT implementation. 
    	
    	Furthermore, we note that the presence of strong laser field breaks the six-fold rotational symmetry of the graphene lattice. For instance, with  $\omega = 4.13$~eV, photoexcitation at two $M'$ points are absent, while excitations at other symmetric $M$ points are observed. This symmetry breaking is caused by presence of linearly polarized laser field along the $y$ direction.

    	It can be explained with a two band model of graphene (see appendix). The excellent agreement on the excitation outcome between the model Hamiltonian and first-principles quantum dynamics simulations justify the validity of our rt-TDDFT algorithm with a vector gauge field. We therefore expect that it is readily applicable to investigate quantum dynamics of a variety of electronic phases such as charge/spin density waves, Mott insulators, valley electronics, and electronic melting in two-dimensional materials and conventional semiconductors. 
    	
    	
    	\begin{figure}
		\centering
		\includegraphics[width=1.0\linewidth]{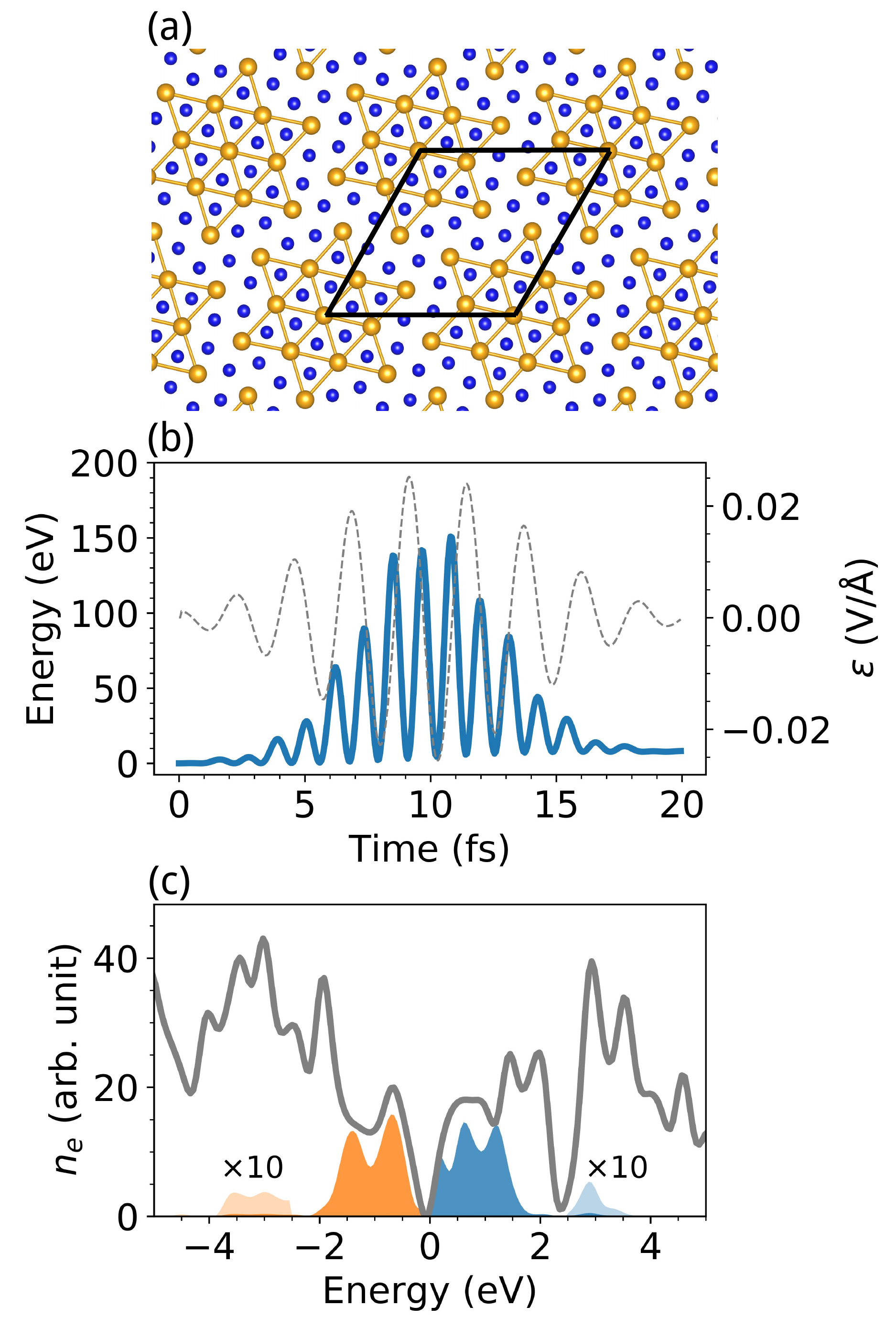}
		\caption{(a) The atomic structure of TaS$_2$. Yellow and blue balls denote the Ta and S atoms, respectively. (b) The excitation energy as a function of time. The grey dash line denotes the field strength of laser pulse. (c) Photoexcitation induced carrier distribution in energy at $t=20$~fs. Yellow and blue fill regions denote the distribution of excited holes and electrons, respectively. The intensity in light yellow and light blue regions are multiplied by ten times. The grey solid line denotes the electronic density of states in ground state.  }
		\label{fig:FigureTaS2}
		\end{figure}

    	To demonstrate the general applicability of the present approach, we tackle photoexcitation induced electron dynamics in a complex material. The layered transition-metal dichalcogenides such as 1T-TaS$_2$ have been widely studied in literature to understand charge density wave (CDW) physics in real materials, whose structure is shown in Fig.~\ref{fig:FigureTaS2}(a). The 1T-TaS$_2$ is a typical quasi two-dimensional CDW material with a pristine lattice constant of 3.36~\AA\ in the undistorted 1T phase. In ground state, the lattice undergoes a structural reconstruction forming a $\sqrt{13}\times\sqrt{13}$ superstructure with star-of-David patterns. Laser induced phase dynamics in 1T-TaS$_2$ has been investigated in recent experiments, where its responses to ultrashort laser pulses play a critical role.      	
    	Here we study the carrier distribution in 1T-TaS$_2$ upon ultrafast laser excitation. As shown in Fig.~\ref{fig:FigureTaS2}(b), the excitation energy strongly oscillates with the field of laser pulse. The excitation energy deposited by the laser pulse is $\sim$12~eV/cell after laser illumination with a photon energy of $\hbar\omega = 1.55$~eV and pulse width of 8~fs. The carrier distribution at 20 fs after the passing of the laser pulse is shown in Fig.~\ref{fig:FigureTaS2}(c). The majority of excited electrons and holes are located at energies ranging from $-$2 to 2 eV near the Fermi level. It indicates that the photoexcitation mainly consists of single-photon processes as well as a minor fraction of two-photon processes (with excited electrons located at $\sim$3 eV and holes at $-3$ eV).

    	\section{Conclusions}
    	In conclusion, we have developed $\mathbf{k}$-resolved rt-TDDFT algorithms using efficient numerical atomic basis. It enables large-scale rt-TDDFT simulations of extended systems including solids, interfaces, and two-dimensional materials with a rather small unit cell, significantly reducing the heavy computational cost of typically rt-TDDFT simulations. Consequently, $\mathbf{k}$-resolved excitation dynamics in periodical crystal materials are observed. The key advantages of this unique approach includes:\\    
	
    	\noindent i) {The $\mathbf{k}$-resolved real-time evolution algorithm introduces the important $\mathbf{k}$-space resolution and a new degree of freedom, which is essential to describe key quantities and important physics in photoexcited condensed matter materials. The use of many $\mathbf{k}$-points with a rather small unit cell also significantly improves the computational efficiency of rt-TDDFT calculations for photoexcitation in solids.\\}
    	
    	\noindent ii)  Different from approaches using real space grids and all-electron full-potential linearized augmented-planewaves, the adoption of numerical atomic basis in the present implementation reduces the number of required basis functions to one-hundredth of its original value, making rt-TDDFT computation of realistic large systems (comprising $\sim$500 atoms and lasting for $\sim$1000 fs) plausible. In addition, with a relatively small real-space cutoff for NAOs, the order-$N$ linear scaling with respect to the system size can be achieved. \\
    	
    	\noindent iii) Both electronic and ionic degree of freedoms are evolved, therefore a complete information on  electronic wavefunctions and ionic movements during real time evolution can be provided for simulations of complex materials and rich phenomena far from equilibrium.\\

    	When applied to study photoexcitation dynamics of a prototypical model material--graphene, the $\mathbf{k}$-resolved algorithm enables the observation of $\mathbf{k}$ selective excitation modes. Three distinct modes are excited, located at different $\mathbf{k}$. In-plance excitation of the Dirac electrons in graphene can be understood by assuming an effective vector field of laser field, via taking into account the angular dependence of optical transition matrix elements. This kind of $\mathbf{k}$ dependent electronic dynamics are ubiquitous in solids. Thus, $\mathbf{k}$-resolved rt-TDDFT algorithm is an important development for investigating ultrafast photoexcitation dynamics and electron-electron scattering, and is expected to be widely used in the future. 
    	
    	\section{acknowledgement}
    	We acknowledge partial financial support from MOST (Grant Nos. 2016YFA0300902 and 2015CB921001), NSFC (Grant Nos. 11774396, 11474328, and 91850120), and CAS (Grant No. XDB07030100).

		\section{APPENDIX: THE TWO-BAND MODEL OF GRAPHENE}
		\beginsupplement

		The ground state Hamiltonian of two-band model of graphene reads,
		\begin{equation}
		H_0(k_x, k_y) = v_F(k_x \sigma_x + k_y \sigma_y),
		\end{equation}
		where $k_x$, $k_y$ is the $\mathbf{k}$ coordinate, $\sigma$ is the Pauli matrices, $v_F$ is the Fermi velocity. $v_F = 1$~eV$\cdot$Bohr for simplification. The units of $k_x$ and $k_y$ are chosen as Bohr$^{-1}$. The energy unit is thus eV. The eigenvalues and eigenvectors are solved as,
		\begin{equation}
		E_0 = -\sqrt{k_x^2 + k_y^2}, 
		\phi_0 = \frac{\sqrt{2}}{2}\left(\begin{array}{c} -1 \\ \frac{k_x + i k_y}{\sqrt{k_x^2 + k_y^2}} \end{array}\right),
		\end{equation}
		\begin{equation}
		E_1 = \sqrt{k_x^2 + k_y^2}, 
		\phi_1 = \frac{\sqrt{2}}{2}\left(\begin{array}{c}1 \\ \frac{k_x + i k_y}{\sqrt{k_x^2 + k_y^2}} \end{array}\right).
		\end{equation}
		Thus, the initial state wavefunction is the ground state $\psi(t=0) = \phi_0$. 
		
		A vector field polarized along $y$ can be introduced as,
		\begin{equation}
		H'(t) = A(t) \sigma_y,
		\end{equation}
		where $A(t)$ is the vector gauge field. The time-dependent Hamiltonian is thus, 
		\begin{equation}
		H(t)  =  H_0 + H'(t).
		\end{equation}
		The wavefunction at time $t$ can be obtained from time-dependent Schr\"{o}dinger equation, 
		\begin{equation}
		\ket{\psi(t)} = \exp[-iH(t)t]\ket{\phi_0},
		\end{equation}
		which can be expanded with $\ket{\phi_0}$ and  $\ket{\phi_1}$ basis,
		\begin{equation}
		\ket{\psi(t)} = c_0(t) \ket{\phi_0}  +  c_1(t) \ket{\phi_1},
		\end{equation}
		where
		\begin{equation}
		c_i(t) = \braket{\phi_i | \psi(t)}
		\end{equation}
		is the time-dependent coefficients. All equations are solved numerically with the \texttt{qutip} package~\cite{Johansson2012, Johansson2013}.
		\begin{figure}
			\centering
			\includegraphics[width=\linewidth]{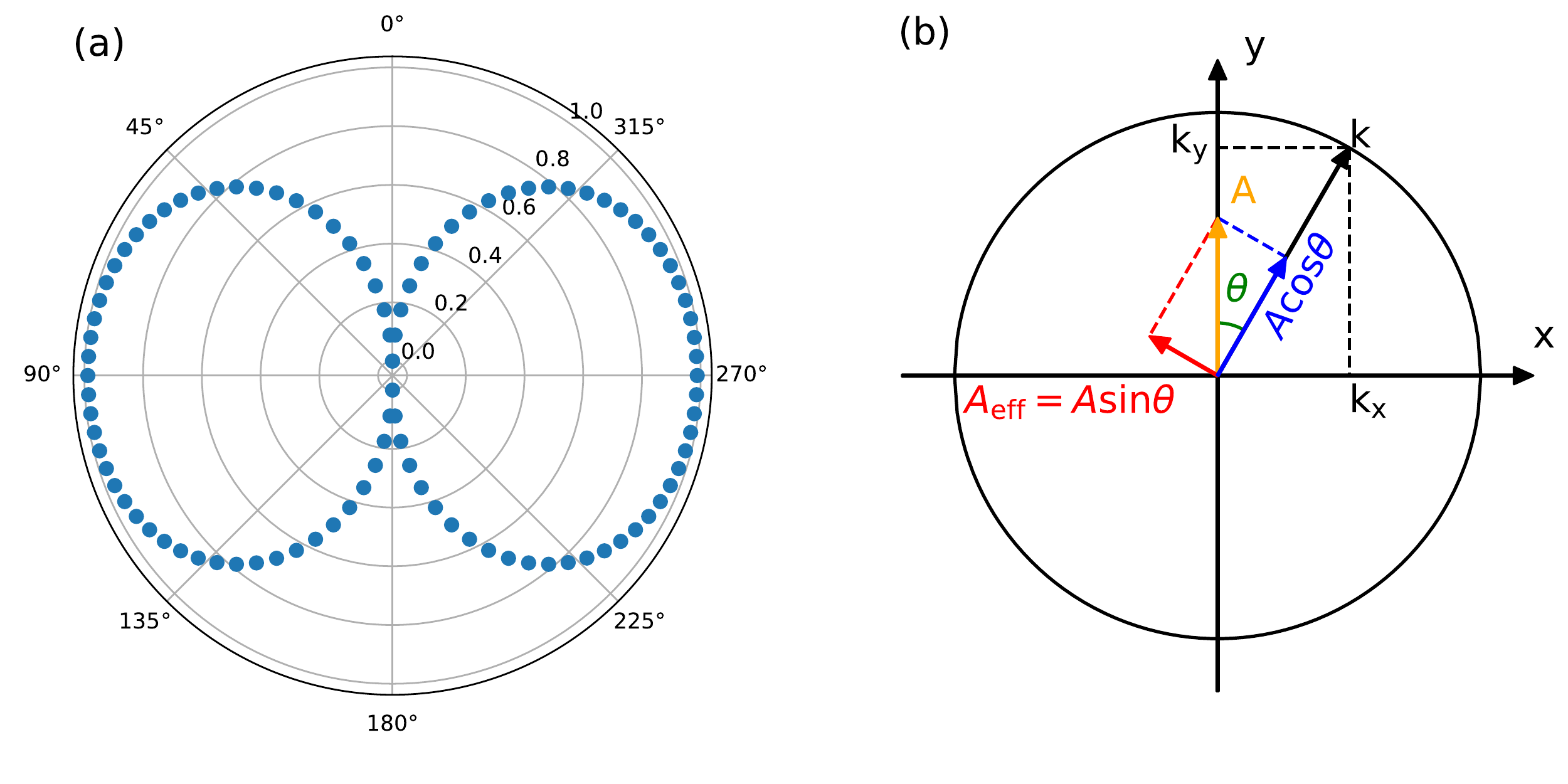}
			\caption{(a) The excited electrons $|c_i(t=50~\mathrm{fs})|^2$ at $\mathbf{k}$ point $(\sin\theta, \cos\theta)$, as a function of the angle $\theta$ between $\mathbf{A}$ and $\mathbf{k}$. (b) Sketch of the effective vector field $A_{eff}$ of a linearly polarized laser field. }
			\label{fig:DemonAngle-figure0}
		\end{figure}

		We can reproduce the symmetry breaking in the distribution of excited electrons in $\mathbf{k}$ space induced by linearly polarized laser. We analyze the coefficients of $|c_2(t)|^2$ with $k_x = \cos\theta, k_y = \sin\theta$, under the vector field $\mathbf{A} = 0.2$~Bohr$^{-1}$, where $\theta$ is the angle between $\mathbf{k}$ and $\mathbf{A}$, as shown in Fig.~\ref{fig:DemonAngle-figure0}. Thus, the energy difference $\Delta E_g(\mathbf{k}) = 2.0$~eV. These $\mathbf{k}$ points are only excited with $\omega = 2.0$~eV, consistent with the results from TDDFT and Fermi's golden rule. 
		
		To explain the origin of the symmetry breaking, 
		the excited electrons at different $\mathbf{k}$ points at the end of laser pulse $|c_2(t=50~\mathrm{fs})|^2$ are shown in Fig.~\ref{fig:DemonAngle-figure0}(a). 
		It suggests that, the effect of linearly polarized laser on point $\mathbf{k}$ is not solely characterized by the $\mathbf{A}$ field, but also related to the angle $\theta$ between $\mathbf{k}$ and $\mathbf{A}$. With $\theta = 0$ and $\pi$, i.e. the $\mathbf{k}$ is parallel/anti-parallel to the $\mathbf{A}$ field, the excitation is fully suppressed, while the excitation is the maximum with $\theta = \pi/2$ and $3\pi/2$. An effective field $A_{\mathrm{eff}} = A\sin\theta$, always perpendicular to the vector $\mathbf{k}$, is thus introduced to induce electronic transitions at  $\mathbf{k}=(k_x = \cos\theta, k_y = \sin\theta)$, as shown in Fig.~\ref{fig:DemonAngle-figure0}(b). It explains the origin of the symmetry breaking in TDDFT simultions (Fig.~\ref{fig:nestingGrapheneAField}). The excitations at $\mathbf{k}$ points are the results of the combined effects of energy match and the angle $\theta$ between the $\mathbf{k} - \mathbf{K}$ and $\mathbf{A}$ field, where $\mathbf{K}$ is the coordinates of the adjacent Dirac point. Since $\mathbf{A}$ field is along $y$, $\sin\theta=0$ for all the $\mathbf{k}$ points with $\mathbf{k} - \mathbf{K}$ parallel to the polarization direction. Thus, this is no effective field to introduce photoexcitations at the two $M'$ points. 
		
		
		%

\end{document}